%% file: Template.tex
\documentclass{article}
\usepackage{spconf,subcaption,cite,amssymb,amsfonts,amsmath,graphicx,hyperref,booktabs,tabularx,svg}
\usepackage[ruled,vlined]{algorithm2e}
\title{UNMIXX: Untangling Highly Correlated Singing Voices Mixtures}
\name{Jihoo Jung, Ji-Hoon Kim, Doyeop Kwak, Junwon Lee, Juhan Nam, Joon Son Chung}
\address{Korea Advanced Institute of Science and Technology, South Korea}
\begin{document}
% \ninept
\maketitle

%\ninept
%
\maketitle
\begingroup
\renewcommand\thefootnote{}
\endgroup
\input{sec/0_abstract}
\input{sec/1_intro}

\input{sec/3_method}
\input{sec/4_exp_result}

\input{sec/5_conclusion}

%\vfill
%\pagebreak

\vfill\pagebreak

\bibliographystyle{IEEEbib}
\bibliography{shortstrings,refs}

\end{document}

%% file: sec/0_abstract.tex
\begin{abstract}
We introduce UNMIXX, a novel framework for multiple singing voices separation (MSVS). While related to speech separation, MSVS faces unique challenges: data scarcity and the highly correlated nature of singing voices mixture. To address these issues, we propose UNMIXX with three key components: (1) musically informed mixing strategy to construct highly correlated, music-like mixtures, (2) cross-source attention that drives representations of two singers apart via reverse attention, and (3) magnitude penalty loss penalizing erroneously assigned interfering energy. UNMIXX not only addresses data scarcity by simulating realistic training data, but also excels at separating highly correlated mixtures through cross-source interactions at both the architectural and loss levels. Our extensive experiments demonstrate that UNMIXX greatly enhances performance, with SDRi gains exceeding 2.2 dB over prior work.

\end{abstract}

%% file: sec/1_intro.tex
\section{Introduction}
\label{sec:intro}
As vocal layering--a technique that stacks multiple vocal tracks--has become a standard practice in contemporary music production, real-world music predominantly features multiple vocal tracks rather than a single line. However, most works on singing voices, such as singing information retrieval~\cite{rosvot2024} and singing voice synthesis~\cite{tcsinger}, assume a single-vocal setting, limiting their applicability to real-world scenarios. Multiple Singing Voices Separation (MSVS) addresses this gap by disentangling individual vocal tracks from complex mixtures, thereby enabling existing methods on singing voices to extend to real-world multi-vocal music.

MSVS is similar to speech separation in that both aim to separate acoustic sources within the same modality-singing voices and speech, respectively. However, MSVS poses greater challenges for two main reasons. First, suitable training datasets are scarce. As shown in Table~\ref{tab:dataset_duration}, unlike the vast speech separation datasets, MSVS datasets contain only about an hour of audio. Second, singing voices exhibit a highly correlated nature. They are often aligned in note onsets and offsets, share harmonic components, contain similar lyrics, and even include segments sung by the same singer. 

\input{table/dataset_stat} Prior studies on MSVS can fall into two categories. One targets choral music \cite{petermann2020deep,gover2020score, sarkar2021vocal}, separating soprano, alto, tenor, and bass from mixtures with four or more voices. The other focuses on pop music, typically with two singers, which is more practical and relevant to real-world use. However, to our knowledge, MedleyVox \cite{jeon2023medleyvox} is the only existing work addressing pop music separation. While it presents an evaluation dataset and baseline study, it still suffers from the two aforementioned challenges of MSVS. First, to compensate for scarce multi-singer training data, they primarily create synthetic mixtures by randomly mixing two monophonic vocals. This approach, however, struggles to capture the complex correlations of real multi-singer mixtures. Second, they largely rely on speech separation frameworks~\cite{luo2019conv, rixen2022sfsrnet}, which are inadequate for disentangling highly correlated mixtures. As a result, remnants of one singer’s voice often remain audible in the other’s output, a phenomenon referred to as \textit{interference}.

In this paper, we propose \textbf{UNMIXX}, a comprehensive framework which mitigates challenges in MSVS with three key components. First, we propose a musically informed mixing strategy that constructs mixtures by combining two songs with strong temporal and harmonic correlation. This produces highly correlated, music-like mixtures resembling real-world multi-singer tracks. Second, we propose cross-source attention, which forces the representations of two singers to diverge via reverse attention~\cite{tao2024seanet}. Third, we propose a magnitude penalty loss that explicitly penalizes spectrogram regions contaminated by interference. Both cross-source attention and magnitude penalty loss enforce mutual exclusivity between outputs through cross-source interactions at architectural and loss levels. This reduces interference and yields cleaner outputs, even in highly correlated mixtures. Experiments validate the effectiveness of UNMIXX, demonstrating consistent improvements on both duet and unison subsets of MedleyVox test set. Audio samples are available here\footnote{\href{https://unmixx.github.io/}{https://unmixx.github.io/}}.

%% file: table/dataset_stat.tex
\begin{table}[t]
    \centering
    \captionsetup{font=small}
    \caption{Public datasets for speech separation and MSVS.}
    \vspace{-2mm}
    \resizebox{0.78\linewidth}{!}{
    \begin{tabular}{l@{\hskip 6mm} l@{\hskip 2mm} c}
        \toprule
        \textbf{Category} & \textbf{Corpus/Dataset} & \textbf{Duration (hours)} \\
        \midrule
        Speech Separation     
            & WSJ0-2mix \cite{hershey2016deep}        & 43 \\
            & Libri2Mix \cite{cosentino2020librimix}  & 292 \\
        \midrule
        MSVS (Choral Music) 
            & jaCappella \cite{nakamura2023jacappella} & 0.9 \\
            & ESMUC Choir \cite{Cuesta2022PhD}        & 0.5 \\
        \midrule
        MSVS (Pop Music) 
            & MedleyVox \cite{jeon2023medleyvox}      & 1.0 \\
        \bottomrule
    \end{tabular}
    }
    \label{tab:dataset_duration}
    \vspace{-4mm}   
\end{table}

%% file: sec/3_method.tex
\section{Overall Architecture}
Our architecture builds upon TIGER~\cite{xu2025tiger}, a lightweight speech separation model, augmented with cross-source attention module in order to identify and suppresses interference. The input mixture is first converted into a time-frequency representation using STFT. The frequency axis is then split into non-uniform sub-bands, with each sub-band projected into a fixed-dimensional space. It is then processed by two key modules--multi-scale selective attention and full-frequency-frame attention (F$^3$A)--applied first along the frequency dimension and then along the time dimension. In our method, F$^3$A incorporates both self-attention and cross-source attention. After repeating this interleaved process eight times, the full-band representation is restored to generate a mask for each singer. These masks are then applied to the input mixture to obtain separated waveforms via inverse STFT.
\section{UNMIXX}
\label{sec:format}
\subsection{Musically Informed Mixing (MIM)}
\input{table/data_mining_figure}
We propose musically informed mixing (MIM) to address the scarcity of multi-singer training data. Similar to~\cite{sep}, rather than randomly mixing two monophonic vocals, MIM selects pairs of songs with strong temporal and harmonic correlation to generate highly correlated mixtures. Temporal alignment serves as a global data mining strategy to create mixtures with better synchronized note on/offsets and harmonic alignment as a local strategy to produce harmonically coherent mixtures.
Figure~\ref{fig:mining} illustrates the pipeline.

\noindent \textbf{Temporal Alignment.} As part of global data mining, we enhance rhythmic consistency by selecting songs with similar tempi and synchronizing segments at downbeat positions. Specifically, before training, we extract beat and downbeat timings from all songs using a recent beat tracking model~\cite{foscarin2024beat}. Each song’s BPM is estimated from the median inter-beat interval, and songs with similar BPM are grouped together. During training, we perform dynamic mixing by randomly choosing a tempo group, sampling two songs from it, and cropping them to a fixed length. Each segment starts at one of downbeat positions rather than an arbitrary point, and the two segments are then mixed to form a training sample.

\noindent \textbf{Harmonic Alignment.} As part of local data mining, we enhance harmonic consistency by constructing each batch exclusively from audio pairs exhibiting strong harmonic correlation. We measure harmonic correlation using harmonic overlap score~\cite{sarkar2021vocal}, which quantifies coinciding partials across the first 16 overtones of the two sources. During training, $B \times M$ candidate pairs are first sampled, where $B$ denotes the batch size and $M$ is a multiplicative factor. Harmonic overlap scores are computed for all pairs, sorted in descending order, and the top $B \times m$ candidates ($m < M$) are retained. From these, a final batch of $B$ pairs is randomly sampled. Here, $m$ is a hyperparameter that controls the degree of harmonic alignment, with smaller values enforcing stronger alignment. In our experiments, we set $M = 16$ and $m = 8$. 
 
\subsection{Cross-Source (CS) Attention}
\label{sec:reverse}

\begin{figure}[t]
  \centering
  \vspace{-2mm}
  \captionsetup{font=small}
  \includegraphics[width=0.85\linewidth]{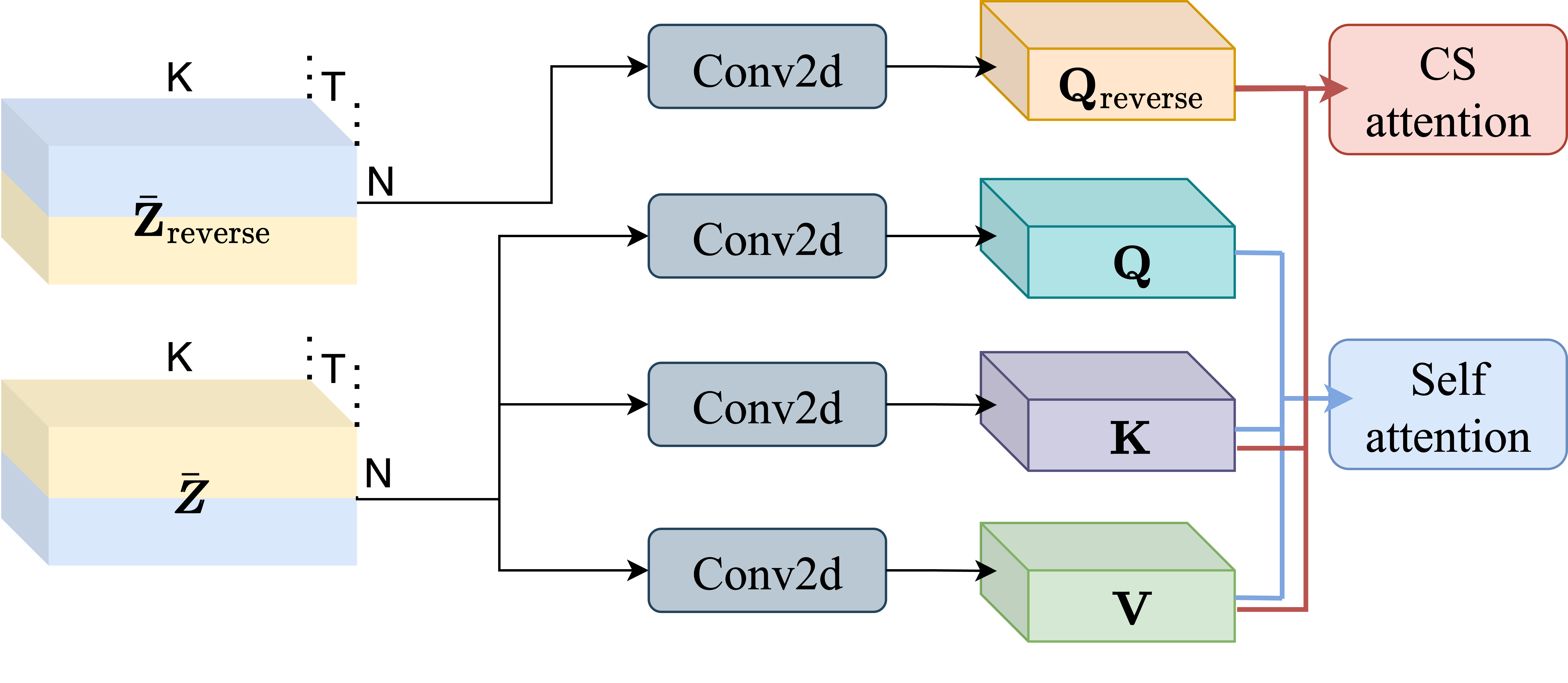}
  \vspace{-2mm}
  \caption{Cross-source attention mechanism.}
  \vspace{-6mm}
  \label{fig:reverse_attention_figure}
\end{figure}  

To promote divergence between the representations of two singers and mitigate interference, we propose cross-source (CS) attention. CS attention first divides the intermediate representation along the channel dimension into two halves, each corresponding to one singer. Then it suppresses regions exhibiting high similarity between the two representations leveraging reverse attention~\cite{tao2024seanet}, a variant of cross-attention in which the logits are negated prior to softmax. 

We extend the self-attention-based F$^3$A module by integrating CS attention. Let the input of the F$^3$A module be $\bar{\pmb{Z}} \in \mathbb{R}^{N \times K \times T}$, where $N$, $K$, and $T$ denote the channel dimension, the number of frequency sub-bands, and the number of frames, respectively. As illustrated in Figure~\ref{fig:reverse_attention_figure}, $\bar{\pmb{Z}}$ is split along the channel dimension $N$, and the front and back halves are swapped to construct a reversed input $\bar{\pmb{Z}}_{\text{reverse}} \in \mathbb{R}^{N \times K \times T}$. To compute reverse attention, a $1 \times 1$ convolution is applied to $\bar{\pmb{Z}}_{\text{reverse}}$ to obtain $\pmb{Q}_{\text{reverse}} \in \mathbb{R}^{(A \times E) \times K \times T}$, while three separate $1 \times 1$ convolutions are applied to $\bar{\pmb{Z}}$ to generate $\pmb{Q} \in \mathbb{R}^{(A \times E) \times K \times T}$, $\pmb{K} \in \mathbb{R}^{(A \times E) \times K \times T}$ and $\pmb{V} \in \mathbb{R}^{(A \times N/A) \times K \times T}$, where $A$ is the number of attention heads and $E$ the embedding dimension per head. CS attention weights are computed as:
\begin{align*}
\pmb{A}_\text{cs} &= 
  \mathrm{Softmax}\!\left( 
    -\frac{\pmb{Q}_{\text{reverse}} \pmb{K}^\top}{\sqrt{E \times T}} 
  \right).
\end{align*}
This formulation down-weights regions of high similarity between the two representations through the negative sign.
At the same time, self-attention weights $\pmb{A}_\text{self}$ are computed from $\pmb{Q}$ and $\pmb{K}$ without negation. 
The final output of the F$^3$A module is the average of self- and CS attention:
$\pmb{O} = \tfrac{1}{2}\big(\pmb{A}_{\text{self}}\pmb{V} + \pmb{A}_{\text{cs}}\pmb{V}\big)$. Here, self-attention preserves the internal consistency of each representation, while CS attention drives the two representations apart. With repeated application of F$^3$A, the representations of the two singers gradually learn to capture mutually exclusive information from the mixture, effectively suppressing interference.

\subsection{Magnitude Penalty Loss}
To further enhance separation quality, we propose a magnitude penalty loss ($\mathcal{L}_{\text{Penalty}}$) that suppresses interference in the predicted magnitude spectrogram. We identify interfering components by comparing each predicted spectrogram with the ground-truth spectrograms of the target and non-target sources. This cross-source constraint enables effective separation even for strongly entangled mixtures.

For each target source $i$, we first construct a binary interference mask $I_i$ by locating 
time-frequency bins that satisfy two conditions: (1) the non-target source $j$'s 
ground-truth magnitude spectrogram $M^{(j)}_{t,f}$ exhibits high energy 
($> \tau_{\max}$), and (2) the target source $i$'s ground-truth magnitude spectrogram
$M^{(i)}_{t,f}$ exhibits low energy ($< \tau_{\min}$). 
Intuitively, $I_i$ 
captures regions that are strongly present in the non-target source but absent 
in the target source, and thus should not appear in the estimated magnitude spectrogram
$\hat{M}^{(i)}$. The magnitude penalty loss is then computed by multiplying this mask $I_i$ with the estimated magnitude spectrogram $\hat{M}^{(i)}$ and normalizing by the number of interfering bins.  In this way, magnitude penalty loss explicitly penalizes $\hat{M}^{(i)}$ by 
capturing undesired energy. This can be formulated as:
\begin{align*}
\mathcal{L}_{\text{Penalty}}
&= \sum_{i=1}^2 \mathbb{E}_{\hat{M}^{(i)}} \left[ 
\frac{\| \hat{M}^{(i)} \odot I_i \|_2^2}
     {\| I_i \|_1 + \epsilon} \right], \\[4pt]
I_i(t,f) &=
\begin{cases}
1 \,, & \text{if } M^{(j)}_{t,f} > \tau_{\max} \ \text{and}\ M^{(i)}_{t,f} < \tau_{\min}, \\
0 \,, & \text{otherwise}.
\end{cases} 
\end{align*}

We combine the proposed magnitude penalty loss with the conventional 
signal-to-noise ratio (SNR) loss ($\mathcal{L}_{\text{SNR}}$) and a magnitude loss 
($\mathcal{L}_{\text{Mag}}$). Magnitude loss is defined as the 
L2 distance between the ground-truth and estimated magnitude spectrograms. The overall training objective is given by
\begin{align*}
\mathcal{L}_{\text{Total}}
&= \mathcal{L}_{\text{SNR}}
+ \lambda_{\text{mag}} \cdot \mathcal{L}_{\text{Mag}}
+ \lambda_{\text{penalty}} \cdot \mathcal{L}_{\text{Penalty}},
\end{align*}
where $\lambda_{\text{mag}}$ and $\lambda_{\text{penalty}}$ are non-negative weights balancing the magnitude and penalty losses. We set $\tau_{\max}=1.0$, $\tau_{\min}=0.5$, $\lambda_{\text{mag}}=0.1$, and $\lambda_{\text{penalty}}=0.02$, with the penalty loss applied after half of the training process. 

%% file: table/data_mining_figure.tex
\begin{figure}[t]
    \centering
    \captionsetup{font=small}
    \begin{subfigure}{1.0\linewidth}
        \centering
        \includegraphics[width=\linewidth]{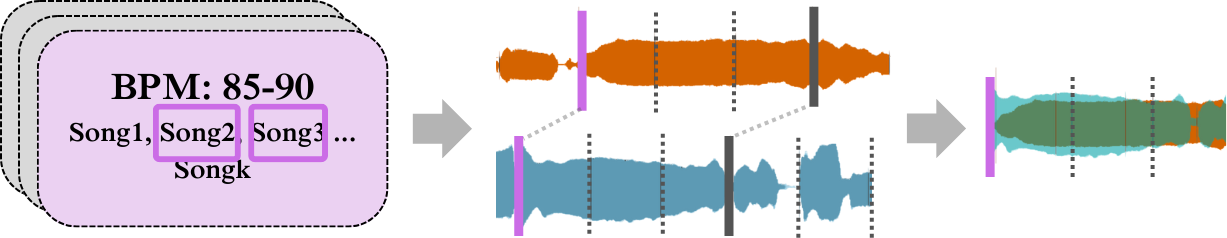}\
        \vspace{-4mm}
        \caption{Illustration of temporal alignment. Two songs are randomly sampled from a tempo group, each cropped to a fixed length starting from one of downbeat positions and then mixed. Dotted, solid, and purple lines denote beats, downbeats, and selected downbeats position, respectively.}

    \end{subfigure}
    \\[2mm]
    \begin{subfigure}{0.9\linewidth}
        \centering
        \includegraphics[width=\linewidth]{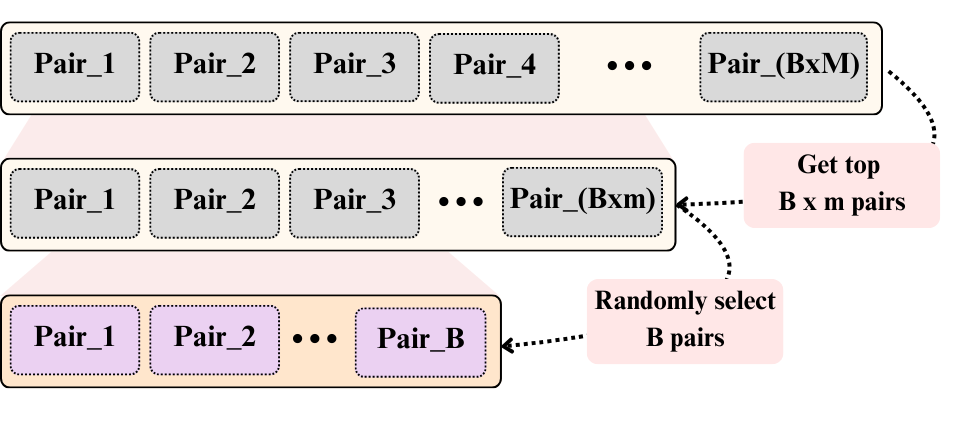}
        \vspace{-6mm}
        \caption{Illustration of Harmonic alignment. The first row shows $B \times M$ audio pairs sorted in descending order by harmonic overlap score. From the top $B \times m$ pairs, the final $B$ pairs are randomly selected.}
    \end{subfigure}
    \vspace{-2mm}
    \caption{Musically informed mixing process.}
    \vspace{-4mm}
    \label{fig:mining}
\end{figure}

%% file: sec/4_exp_result.tex
\section{Experimental Setup}
\label{sec:pagestyle}
\subsection{Datasets and Training Details}
We train on around 400 hours of audio from 9 monophonic singing datasets~\cite{choi2020children, duan2013nus, wilkins2018vocalset, jsut_song, tamaru2020jvs, schulze2021phoneme, k_multitimbre, k_multitisinger, musdb18-hq}. 
For evaluation, we use the \textit{unison} and \textit{duet} subsets of MedleyVox evaluation dataset. 
The \textit{unison} subset consists of mixtures of two voices with identical or octave-shifted melodies, same note on/offsets and lyrics, while the \textit{duet} subset contains mixtures differing in melodies, note on/offsets, or lyrics. Both subsets contain mixtures of either two different singers or two parts sung by the same singer. 
Notably, 55\% of duet mixtures and 97\% of unison mixtures involve multiple parts sung by the same singer. 
All audio is resampled to 24 kHz. We compute STFTs with a 960-sample window, 240-sample hop, and 960-point FFT, and apply power-law compression to magnitude spectrograms to reduce dynamic range. Models are trained with a batch size of 8 using Adam optimizer at a 0.001 learning rate decayed by validation performance. Training runs for up to 500k steps, with early stopping after 60k steps of no improvement.

\subsection{Evaluation Metrices}
Following prior work~\cite{jeon2023medleyvox}, we use SDRi and SI-SDRi as evaluation metrics. However, since these metrics often underestimate quality in same-singer mixtures, we propose permutation-invariant segmental SNR (PSSNR) and its hybrid variant HSSNR as auxiliary metrics tailored for MSVS. 

SDRi and SI-SDRi often yield misleadingly low scores for same-singer mixtures, as they heavily penalize singer assignment changes. Such cases arise when the model swaps singers (e.g., $S_1$ and $S_2$) across segments ($S_1S_2S_1$ and $S_2S_1S_2$) instead of producing consistent outputs ($S_1S_1S_1$ and $S_2S_2S_2$). While such penalties are appropriate in speech separation with distinct speakers, they are unnecessary in singings when a single singer performs multiple parts.

To address this, we propose PSSNR, which is identical to SSNR except that it recomputes the optimal permutation for each segment. PSSNR focuses solely on separation quality at segment level, ignoring assignment consistency across segments. To provide a unified score across different- and same-singer cases, we define HSSNR as the average of SSNR for different-singer cases and PSSNR for same-singer cases. In this way, HSSNR reflects necessary penalties from singer assignment changes while excluding redundant ones.

\input{table/synthetic_sdr}
Table~\ref{tab:synthetic_results} presents the limitations of SDRi and SI-SDRi under singer assignment changes on the same-singer unison subset, while PSSNR mitigates them. We simulate assignment changes by randomly swapping segments between two ground-truth signals at specified ratios. This induces assignment changes but is perceptually natural, as both parts are sung by the same singer and separation remains perfect. Nevertheless, SDRi and SI-SDRi drop sharply; beyond 30\% swapping, SI-SDRi even becomes negative, implying that input mixture scores higher than the perfectly separated signals. SSNR shows a similar trend, whereas PSSNR remains stable. This shows that, unlike SDRi, SI-SDRi, and SSNR--which rely on a fixed global permutation--PSSNR removes spurious penalties and thus more reliable in same-singer mixtures.
\section{Experimental Results}

\subsection{Quality Comparison}
\input{table/main_exp}

Table~\ref{tab:main_results} shows the performance of UNMIXX alongside two baselines: MedleyVox~\cite{jeon2023medleyvox} and TIGER~\cite{xu2025tiger} trained on the same datasets as MedleyVox--i.e., our training dataset plus the speech dataset \cite{cosentino2020librimix}.
Across all metrics, UNMIXX consistently outperforms the baselines. Against MedleyVox, UNMIXX delivers substantial SDRi gains of +2.42~dB (duet) and +2.26~dB (unison). Compared to TIGER, it also achieves marked gains of +0.94~dB (duet) and +1.20~dB (unison), with only a marginal parameter increase. Moreover, UNMIXX shows clear improvements in HSSNR, confirming that such substantial SDRi and SI-SDRi gains result from genuine separation quality rather than favorable singer assignments.

\subsection{Ablation Studies}
\label{sec:ablation}

\input{table/ablation}
\input{figure/penaly_loss}
We verify the effectiveness of each  component of UNMIXX through ablation studies, as reported in Table~\ref{tab:ablation_data_mining}. Note that each component is individually added to the TIGER baseline.

The second block evaluates MIM. Excluding speech mixture from training data improves separation quality in both subsets. Building on this, temporal alignment and weak harmonic alignment ($m{=}12$) further enhance performance. Stronger harmonic alignment (smaller $m$) boosts unison but degrades duet performance. This is because excessive alignment produces overly correlated samples and reduces training data diversity, which can harm duet performance as they are inherently less correlated than unison.

The third block evaluates CS attention, which yields consistent gains on both subsets, with particularly large improvements on duet subset. The fourth block examines magnitude-based objectives. Incorporating magnitude loss improves performance, and adding the magnitude penalty loss provides further gains--modest in duet but more pronounced in unison. This is because the duet subset is relatively easy to separate, leaving fewer regions for the penalty to act on, whereas the more challenging unison subset benefits more from such suppression. HSSNR shows a notable 0.51 dB increase in unison, highlighting the effectiveness of the magnitude penalty loss in eliminating residual interference and enhancing separation fidelity. We also visualize the output spectrograms to further demonstrate effectiveness of magnitude penalty loss.
As shown in Fig.~\ref{fig:penalty_loss}, adding the magnitude penalty loss yields clean spectrograms close to the ground truth, whereas the magnitude loss alone results in noisy spectrograms.

%% file: table/synthetic_sdr.tex
\begin{table}[t]
    \centering
    \captionsetup{font=small}
    \caption{Metric values on the unison subset in the same-singer case, obtained by swapping ground-truth signals at specified ratio.}
    \vspace{-2mm}
    \resizebox{0.75\linewidth}{!}{
    \begin{tabular}{l@{\hskip 2mm} c@{\hskip 2mm} c@{\hskip 2mm} c@{\hskip 2mm} c}
        \toprule
        \textbf{Swap ratio(\%)} & \textbf{SDRi~($\uparrow$)} & \textbf{SI-SDRi~($\uparrow$)} & \textbf{SSNR~($\uparrow$)} & \textbf{PSSNR~($\uparrow$)} \\
        
        \midrule
        10 &  7.36  &  6.95  &  31.04  &  34.74 \\
        20 &  3.74  &  2.95  &  26.86  &  34.74 \\
        30 &  1.02  & -0.07  &  22.74  &  34.75 \\
        40 & -0.52  & -1.96  &  17.65  & 34.75 \\
        50 & -1.04  & -2.41  &  15.83  & 34.74 \\
        \bottomrule
    \end{tabular}
    }
    \label{tab:synthetic_results}
    \vspace{-2mm}
\end{table}

%% file: table/main_exp.tex
\begin{table}[t]
    \centering
    \captionsetup{font=small}
    \caption{Separation performance on duet and unison subsets of MedleyVox evaluation dataset. TIGER* denotes the TIGER trained on the same training data as used in the MedleyVox experiments.}
    \vspace{-2mm}
    \resizebox{0.99\linewidth}{!}{
    \begin{tabular}{l@{\hskip 2mm} c@{\hskip 2mm} c@{\hskip 2mm}c@{\hskip 2mm}c@{\hskip 2mm}c@{\hskip 2mm}c@{\hskip 2mm}c}
        \toprule
        \textbf{Method} & \textbf{\#params} 
        & \multicolumn{3}{c}{\textbf{Duet}} 
        & \multicolumn{3}{c}{\textbf{Unison}} \\
        \cmidrule(lr){3-5} \cmidrule(lr){6-8}
        & & SDRi & SI-SDRi & HSSNR & SDRi & SI-SDRi & HSSNR \\
        \midrule
        MedleyVox        & 5M   & 15.10 & 14.20 &   13.33   & 4.90 & 4.40 &   7.65   \\
        TIGER$^*$     & 947k & 16.58 & 15.52 & 15.14  & 5.96 & 5.31 &  9.86  \\
        \textbf{UNMIXX}& 951k & \textbf{17.52} & \textbf{16.47} & \textbf{15.96} 
                                   & \textbf{7.16} & \textbf{6.58} & \textbf{10.50} \\
        \bottomrule
    \end{tabular}
    }
    \vspace{-6mm}
    \label{tab:main_results}
\end{table}

%% file: table/ablation.tex
\begin{table}[t]
    \centering
    \vspace{2mm}
    \captionsetup{font=small}
    \caption{Ablation studies on the duet and unison subsets of MedleyVox evaluation dataset. Each proposed component was added to the baseline individually.  \underline{Underline} highlights the best score within each block. \textbf{Bold} marks the overall best result across the entire table.}
    \vspace{-2mm}
    \resizebox{0.99\linewidth}{!}{
    \begin{tabular}{c l c@{\hskip 1mm}c@{\hskip 2mm}c@{\hskip 2mm}c@{\hskip 2mm}c@{\hskip 2mm}c}
        \toprule
         & \textbf{Method}
         & \multicolumn{3}{c}{\textbf{Duet}}
         & \multicolumn{3}{c}{\textbf{Unison}} \\
        \cmidrule(lr){3-5} \cmidrule(lr){6-8}
         & & SDRi & SI-SDRi & HSSNR & SDRi & SI-SDRi & HSSNR \\
        \midrule
        (1) & TIGER$^*$ & 16.58 & 15.52 & 15.14 & 5.96 & 5.31 & 9.86 \\
        \midrule \midrule
        (2) & - Speech dataset      & 16.57 & 15.46 &   15.71   & 6.54 & 5.90 &   9.89   \\
            & \quad + MIM ($m{=}12$) & \underline{17.11} & \underline{16.05} &   15.43   & 7.03 & 6.43 &   10.06   \\
            & \quad   + MIM ($m{=}8$)  & 16.79 & 15.75 & \underline{15.83} & \textbf{7.31} & \textbf{6.68} &   \textbf{10.72}   \\
            & \quad   + MIM ($m{=}4$)  & 16.09 & 14.96 & 14.58 & 7.12 & 6.48 &   9.50   \\
        \midrule \midrule
        (3) & + CS attention   & \textbf{18.01} & \textbf{17.00} & \textbf{16.02}
                               & \underline{6.17} & \underline{5.54} & \underline{10.06} \\
        \midrule \midrule
        (4) & + Mag loss            & 16.66 & 15.60 &   \underline{15.63}   & 6.26 & 5.71 & 9.29 \\
            & + Mag, Penalty loss   & \underline{16.68} & \underline{15.61} &  15.50
                               & \underline{6.44} & \underline{5.83} & \underline{9.89} \\
        \bottomrule
    \end{tabular}
    }
    \vspace{-2mm}
    \label{tab:ablation_data_mining}
\end{table}

%% file: figure/penaly_loss.tex
\begin{figure}[t]
  \centering
  \vspace{1mm}
  \captionsetup{font=small}
  \begin{subfigure}[b]{0.32\linewidth}
    \centering
    \includegraphics[width=\linewidth,height=0.22\textheight,keepaspectratio]{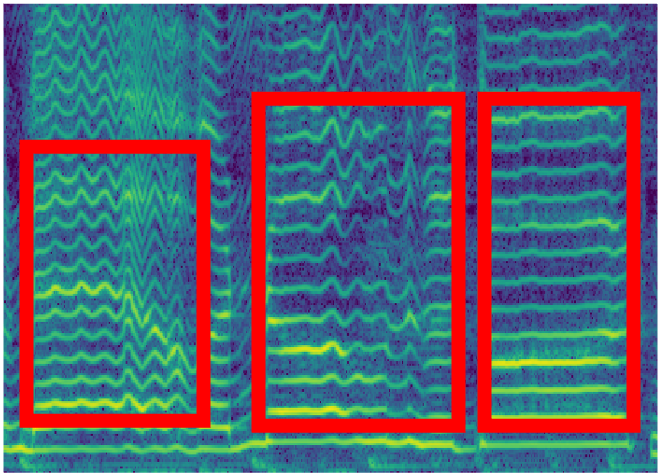}
    \caption{Ground truth}\label{fig:img1}
  \end{subfigure}\hfill
  \begin{subfigure}[b]{0.32\linewidth}
    \centering
    \includegraphics[width=\linewidth,height=0.22\textheight,keepaspectratio]{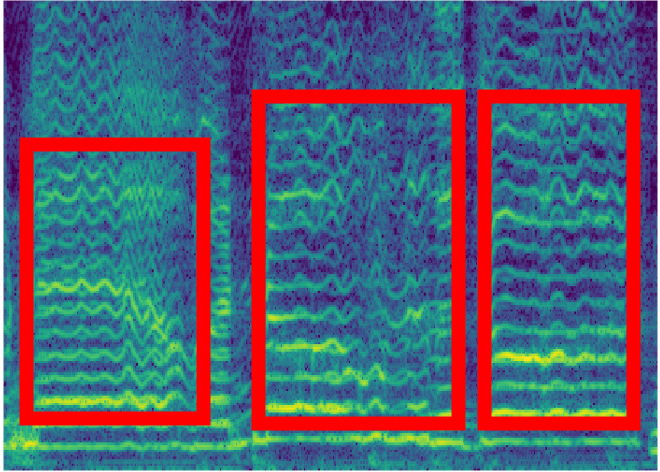}
    \caption{+Mag loss}\label{fig:img2}
  \end{subfigure}\hfill
  \begin{subfigure}[b]{0.32\linewidth}
    \centering
    \includegraphics[width=\linewidth,height=0.22\textheight,keepaspectratio]{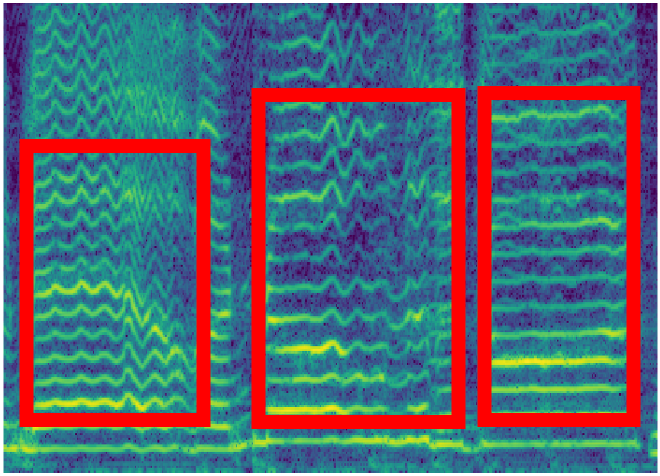}
    \caption{+Mag, Penalty loss}\label{fig:img3}
  \end{subfigure}
  \vspace{-2mm}
 \caption{Comparison of spectrograms using different objectives.}
  \label{fig:penalty_loss}
  \vspace{-6mm}
\end{figure}

%% file: sec/5_conclusion.tex
\section{Conclusion}
\label{sec:majhead}
We propose UNMIXX, a comprehensive MSVS framework for separating individual vocal tracks from complex mixtures. We address the inherent challenges of MSVS with three key components--musically informed mixing, cross-source attention, and magnitude penalty loss--achieving notable gains over prior work. We verify the effectiveness of each component through extensive ablation studies with newly proposed evaluation metrics--HSSNR--tailored for MSVS.